# A STATISTICAL OVERVIEW ON DATA PRIVACY [†]

Fang Liu, Ph. D.*

## Abstract

*The eruption of big data with the increasing collection and processing of vast volumes and variety of data have led to breakthrough discoveries and innovation in science, engineering, medicine, commerce, criminal justice, and national security that would not have been possible in the past. While there are many benefits to the collection and usage of big data, there are also growing concerns among the general public on what personal information is collected and how it is used. In addition to legal policies and regulations, technological tools and statistical strategies also exist to promote and safeguard individual privacy, while releasing and sharing useful population-level information. In this overview, I introduce some of these approaches, as well as the existing challenges and opportunities in statistical data privacy research and applications to better meet the practical needs of privacy protection and information sharing.*

## Introduction

Today, digitization touches every part of our lives, affecting how we work and live. Emerging technologies, such as artificial intelligence, high performance computing, cloud storage, and computing, help to accelerate the digital transformation. As technology capabilities continue to expand, there is also a growing concern in the public around increasing collection, storage, dissemination, and processing of personal information. According to the Pew Research Center, roughly sixty percent of U.S. adults do not think it is possible to go through daily life without their personal data collected by either companies or the government, and over eighty percent are concerned about what is being done with their data.[1]

The issue of data privacy is not new and can be dated back to 1890, when two U.S. lawyers, Samuel D. Warren and Louis D. Brandeis, wrote *The Right to Privacy* to declare a right to privacy in the United States.[2] Fast forward to today, not only are there legal policies and regulations, but also administrative controls, and computer and statistical strategies to promote and safeguard privacy. For example, the European Union enforced the implementation of the General Data Protection Regulation (GDPR) on May 25, 2018; the U.S. privacy laws Health Insurance Portability and Accountability Act (HIPAA) and Children's Online Privacy Protection Act (COPPA) protect personal information of patients since 1996 and of children under thirteen since 1998, respectively; and China has recently adopted the National Standard of Information Security Technology—Personal Information Security Specification.[3] Technology and methodologies have also been in development to protect privacy, such as access control (e.g., password and electronic gatekeepers for remote access to computer databases), destruction (deletion of the data that are no longer needed), digital passport, data anonymization and pseudonymization, and encryption, among others.

Anonymization and pseudonymization relate the most to the statistical strategies for protecting privacy and are often regarded as a must when performing scientific or statistical research. Specifically, anonymization and pseudonymization consist of stripping individual identifying information or injecting noises into the original data before releasing the data and information to the public. This greatly reduces the identification risk of individuals and the disclosure risk of attribute information. On the other hand, due to the perturbation used in the data to achieve anonymization and pseudonymization, there will be unavoidable information loss and the released information to the public will not be as accurate or precise as the original information. Ideally, a data perturbation approach should maximize the protection of respondents in a data set, while minimizing the information loss due to the perturbation. However, in reality, the more protection there is, the less useful the released data are. Figure 1 illustrates how information starts to get lost and privacy increases as more noises and perturbations are injected into the original data, which are the images of human faces. An optimal or near-optimal balance between the two extremes—maximal information and no privacy for individuals versus zero information and maximal privacy for individuals—is often the goal when developing strategies for data release.

---

† To appear in *Notre Dame Journal of Law, Ethics & Public Policy* (2020), Volume 34 Issue 2.

* Applied and Computational Mathematics and Statistics, University of Notre Dame, Notre Dame, IN 46530. The research is partially supported by the NSF Grant #1717417.

1. Brooke Auxier et al., Pew Research Ctr., Americans and Privacy: Concerned, Confused and Feeling Lack of Control Over Their Personal Information (2019), https://www.pewresearch.org/internet/wp-content/uploads/sites/9/2019/11/Pew-Research-Center_PI_2019.11.15_Privacy_FINAL.pdf.

2. Samuel D. Warren & Louis D. Brandeis, *The Right to Privacy*, 4 Harv. L. Rev. 193 (1890).

3. DLA Piper, Data Protection Laws of the World (2019), https://www.dlapiperdataprotection.com/system/modules/za.co.heliosdesign.dla.lotw.data_protection/functions/handbook.pdf?country-1=CN.

**Figure 1**: Data utility decreases and privacy increases as perturbation increases

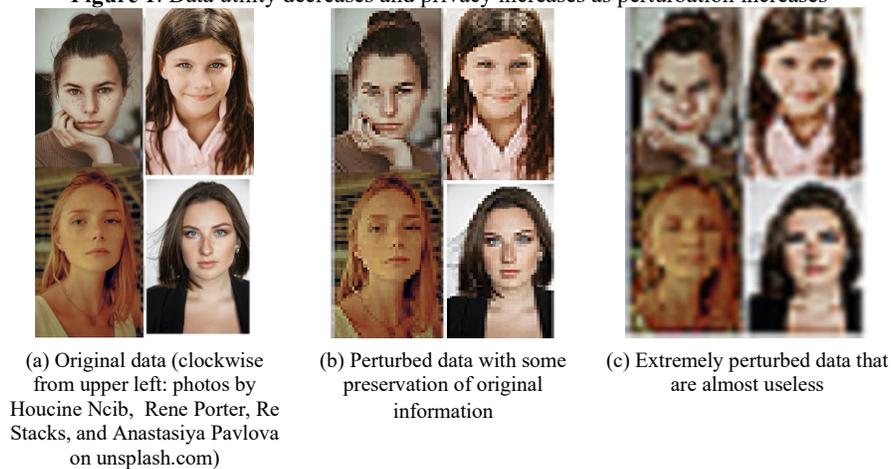

(a) Original data (clockwise from upper left: photos by Houcine Ncib, Rene Porter, Re Stacks, and Anastasiya Pavlova on unsplash.com)

(b) Perturbed data with some preservation of original information

(c) Extremely perturbed data that are almost useless

In what follows, I will first provide an overview on the assessment of privacy risk in released data in Section I, and then present some approaches in Section II to assess utility of the data if they are processed for privacy protection purposes before being released to the public. I will use Section III to introduce some traditional and state-of-the-art methods and techniques for perturbing the original data to lower the privacy risk while maintaining some level of utility upon release. Discussions and concluding remarks are provided in Section IV.

## I. Privacy Risk Measurement

One of the research topics in data privacy is to measure the privacy risk level in the released data and whether the individual privacy can be safeguarded when releasing information. There are two main types of privacy risk: re-identification risk and attribute disclosure risk. The former occurs when a data intruder is able to recognize an individual from the released data, and the latter refers to when a data intruder can figure out the value on an attribute of his target, even without precisely identifying the records from the released data.

Privacy risk measures can be roughly grouped in two categories: relative privacy risk measures and absolute privacy risk measures.[4] The former is also referred to by many as "differential privacy" (DP)[5] or "formal privacy".[6] In relative privacy risk measures, there is a pre-set privacy parameter that upper bounds the *additional* leakage of privacy with each release of information. Data perturbation mechanisms build in the framework of DP and are designed to achieve the pre-specified privacy level. Since the privacy risk control is guaranteed upfront, there is often no need to perform a post-hoc assessment of the privacy risk in the released perturbed data, and one may focus on achieving the highest utility possible for the released information when developing a differentially private perturbation mechanism. In contrast, absolute privacy risk measures focus on quantifying the totality of the privacy risk in released data. In the rest of this Section, I will first examine the absolute privacy risk, and then present the DP concept.

### I.1. Absolute Privacy Risk

#### I.1.1. *k*-Anonymity, *l*-Diversity, and *t*-Closeness

The concept of *k*-anonymity was first introduced by Samarati and Sweeney in 1998.[7] A release of data is said to have the *k*-anonymity property if the information for each person contained in the release cannot be distinguished from at least $k − l$ individuals in the released data. For example, if $k = 2$, then there is no unique record in the released data and each record looks the same to at least one other records. On the other hand, achieving *k*-anonymity

---

exactly can be computationally infeasible when the number of attributes are large; but approximate methods often yield effective results.[8] Though $k$-anonymity is a promising concept for privacy protection, data of $k$-anonymity are still susceptible to many attacks, especially when data attackers possess some background knowledge.[9] In addition, $k$-anonymity does not include any randomization, attackers can still make inferences that may harm individuals. Finally, it is not effective for high-dimensional data (there are many attributes in the data, or the so-called "large-$p$" problem in statistics).

The $l$-diversity is an extension of the $k$-anonymity model.[10] The $l$-diversity model handles some of the weaknesses of $k$-anonymity. For example, when the sensitive values within a group are homogenous, the $l$-diversity model promotes the intra-group diversity on a sensitive attribute. Instead of every individual having the same value on the attribute, there are at least $l$ "well represented" values for the sensitive attribute.[11] "Well represented" in this context can be interpreted in three ways: (1) at least $l$ distinct values; (2) entropy $l$-diversity; or (3) recursive $(c - l)$-diversity that ensures that the most common value does not appear too often while less-common values are ensured to not appear too infrequently. Similar to the $k$-anonymity, the $l$-diversity model does not scale well with the number of attributes. In addition, it might still leak sensitive information when the "diversity" is defined at a granular level and those diverse values may still be semantically close. For example, a data intruder can still conclude an individual has a stomach disease if a data set to which the individual belongs only lists three different stomach diseases. The $t$-closeness is a further refinement of the $l$-diversity that deals with some of its drawbacks.[12] The threshold $t$ gives an upper bound on the difference between the distributions of the sensitive attribute values within an anonymized group as compared to the global distribution of values for numeric attributes.[13]

### I.1.2.    Other Approaches

All the three privacy risk concepts discussed in Section I.1.1 have pre-specified privacy parameters. Once the parameters are set, a protocol or an algorithm is developed and implemented to achieve the pre-set privacy level for the released data; and there is often no need for further privacy risk assessment after the data are released. Besides the parameterized privacy risk models, there exist other privacy risk assessment approaches that are not necessarily associated with a privacy parameter, but rather aim at quantifying re-identification risk or attribute disclosure risk in the released data, often by imposing assumptions on the external knowledge and behaviors of data intruders. These approaches can be regarded as the post-hoc privacy risk assessment as they are carried out upon receiving the data and depend on the approaches used to perform the risk assessment, the assumptions imposed, and sometimes the type of released data.

One of these approaches is through the record-linkage technique that links the released data with external knowledge that a data intruder possesses to re-identify his target. Depending on the assumption on the external knowledge, the re-identification risk varies. In that sense, the approach is, to a degree, ad-hoc and non-robust. While one can always adopt the worst-case scenario approach by assuming the intruder has maximal external knowledge he could use to link to the released data, the meaning of "worst-case" and its supposed guarantee only pertain to the status quo and is not future-proof. In other words, the "worst-case" scenarios may no longer hold if the data intruder obtains more information in the future, and the data perturbation methods built upon it would no longer maintain the originally desired confidentiality levels.

The other privacy risk approach quantifies the probability that a sample unique is also a population unique. A "sample unique" refers to the case that there is only one individual in the sample data that possesses certain values of the attributes; and a "population unique" means there is only one individual in the population that possesses certain values of the attributes. When a sample unique is a population unique, the re-identification risk of that individual is maximal. If the data intruder knows his target is in the released data and has information on all the pseudo-identifiers for his target, and if that target is a sample unique, then that target will be identified even if the target is not a population unique. Therefore, the existence of sample unique per se imposes privacy risk. The good news is that while every individual in the population is unique, sample data are finite in terms of the number of attributes. Therefore, it is likely that some individuals share the same set of attribute values in the sample, especially when the sample size (the total number of individuals) is large.

For the approaches without formal privacy parameters but focusing on the post-hoc privacy risk assessment, the privacy risk will vary by perturbation mechanisms. Some approaches will lead to better privacy protection than the others per either the record-linkage assessment, the population unique probability quantification, or other risk assessment metrics. In other words, not only will different approaches lead to different utility for the perturbed data, but also their privacy guarantees are different. This will make the comparison of perturbation approaches harder than in the setting of formal privacy, where every perturbation approach is under the same privacy guarantee, and those that offer higher utility are preferred.

## I.2. Differential Privacy

Differential privacy (DP), a concept popularized in the theoretical computer science community,[14] provides strong privacy guarantee in mathematical terms without making assumptions about the background knowledge of data intruders. It is also immune to post-processing and is future-proof, meaning the achieved privacy level for the individuals in the released data will remain the same regardless of what post-processing procedure is applied to the data and what future information the data intruder will have about his target. DP is claimed to provably thwart any privacy attack, including the re-identification and attribute disclosure risks.[15] DP is the state-of-the-art concept in data privacy research and has also started to influence how practitioners (industry and government) collect and release data. In what follows, I will provide more details on the concept of DP.

### I.2.1. The Classical Definition

In brief, an $\epsilon$-differentially private mechanism perturbs a statistic (a quantity calculated from a data set) to satisfy the following condition: when the statistic is calculated from two neighboring data sets that differ by one record, the ratio of the probabilities of that statistic taking the same value (any) based on the two sets, is bounded between $(e^{-\epsilon}, e^{\epsilon})$. The perturbed statistic via an $\epsilon$-differentially private mechanism is often referred to as the sanitized statistic. In layman's terms, DP means that the chance an individual will be identified based on the sanitized statistic is low if $\epsilon$ is small since the statistic would be about the same with or without the individual in the database. Inversely, if $\epsilon$ is larger, then the sanitized statistic will be different when an individual is absent or present in the data, leading to a higher chance of that individual being identified and his attribute values being disclosed from the released sanitized results.

To achieve $\epsilon$-DP, noises will be injected to the original statistics to obtain sanitized statistics to release to the public. The higher the requirement on privacy, the smaller the $\epsilon$, the more noises are needed to perturb the original results to achieve $\epsilon$-DP, and the less useful the released sanitized results will be. The desire to achieve a higher level of utility has motivated the work on relaxing the original DP definition, such as the approximate differential privacy,[16] the probabilistic differential privacy,[17] the random differential privacy,[18] and the concentrated differential privacy.[19] For all these relaxed versions of the classical DP, there is at least one more parameter in addition to $\epsilon$ that governs the amount of relaxation.

As can be observed from the discussion above, $\epsilon$ is a critical parameter when it comes to the practical implementation, as it quantifies the privacy level of the sanitized results and affects the usefulness of the released information. Regarding what value of $\epsilon$ is appropriate or acceptable for practical use, Dwork states that the choice of $\epsilon$ is a social question but suggests small values like 0.01, 0.1, or even as large as ln2 or ln3.[20] Lee and Clifton

---

suggest a formula to calculate $\epsilon$ if the goal is to hide any individual's presence (or absence) in the database.[21] Abowd and Schmutte address the question from the economic perspective by accounting for the public-good properties of privacy loss and data utility, and define the optimal choice of $\epsilon$ by formulating a social planner's problem.[22] In the numerical examples published in the literature on DP, many examine a range of $\epsilon$, but often curb the maximum $\epsilon$ at 1.[23] An application of DP to the U.S. Census Bureau's OnTheMap data (commuting patterns of the U.S. population) uses $(\epsilon = 8.6, \delta = 10^{-5})$-probabilistic DP. In summary, there are many factors that affect the choice of $\epsilon$—the social perception of privacy, the sensitivity level of the information to be released, and the desired utility, among others. The choice of $\epsilon$ is still an active research area and might take some time before a consensus, if possible, can be reached.

### I.2.2. Extension of Classical Differential Privacy

*Local Differential Privacy*

One of the population extensions of the classical DP is the so-called local DP.[24] Formally, local DP bounds the likelihood ratio of obtaining the same response through a randomization mechanism, when the true response takes any two different values, below $e^{\epsilon}$. In other words, it suggests that the likelihood that a subject produces a randomized response is close to a constant (how close depends on the value of $\epsilon$) regardless what his true response is. The concept of local DP captures a type of plausible deniability, which means people have the ability to deny knowledge because of a lack of evidence that can confirm their participation, even if they were personally involved in, or at least willfully ignorant of, the actions.[25]

Randomized response, a mechanism proposed in 1965 for data collection with privacy protection,[26] can be regarded as a form of the local DP. There are several ways to conduct randomized response. For example, suppose a subject is asked a sensitive and private question, say, on whether he had ever stolen. Before answering the question, the subject is instructed to flip a coin, and answer truthfully if the coin lands on tails or flip a second coin if it lands on heads. If the second coin comes up heads, then the subject is to answer "Yes," or answer "No" if lands on tails. If the coin is unbiased and the true population proportion of ever stealing is $\theta$, then the expected probability of getting a "Yes" response via randomized response is $0.25 + 0.50\,\theta$ and that of getting a "No" response is $0.25 + 0.50(1 - \theta)$. Interpreted in the context of local DP, it means that if the subject's response is "Yes," then he can claim that is because the coin landed heads up and his privacy is protected. It is easy to prove that the likelihood ratio of having a "Yes" response through the two-time coin flipping randomized response mechanism when the true response is "Yes" versus when the true response is "No" is bounded below 3 (that is, $\epsilon = ln\,(3)$). Data obtained through the randomized response mechanism still allow us to find an unbiased estimate of $\theta$, which is a population-level parameter of interest, even without knowing the true response from each subject or who answers the questions truthfully.

*Location Privacy with Geo-Indistinguishability*

It is very common nowadays that untrusted servers collect massive information on users' location with the increased popularity of mobile devices. Location information is sensitive; no one wants to be followed or "spied on" (even virtually), especially by someone that they do not trust or know. Andrés et al. propose the concept of geo-indistinguishability, a formal notion of location privacy that extends the classical DP framework.[27] Geo-indistinguishability formalizes the intuitive notion that the more precise the location information is about the user, the more sensitive that information is and the more privacy concern there is for the user. The precision of the location information and the privacy protection level can be characterized by the radius $r$ of the neighborhood around the user's true location–the smaller $r$ is, the less privacy protection there is. Geo-indistinguishability states that the user enjoys $(\epsilon \times r)$-privacy, if the randomization mechanism draws a random location from the neighborhood of radius $r$ around the true location. $\epsilon$ corresponds to the level of privacy for one unit of distance, and $\epsilon \times r$ quantifies the overall privacy level for the randomized mechanism. Figure 2 shows that the privacy protection level increases as the radius $r$ of the neighborhood around the user's true location increases.

**Figure 2**: Privacy protection level increases with the radius of the neighborhood around the user's true location per geo-indistinguishability[28]

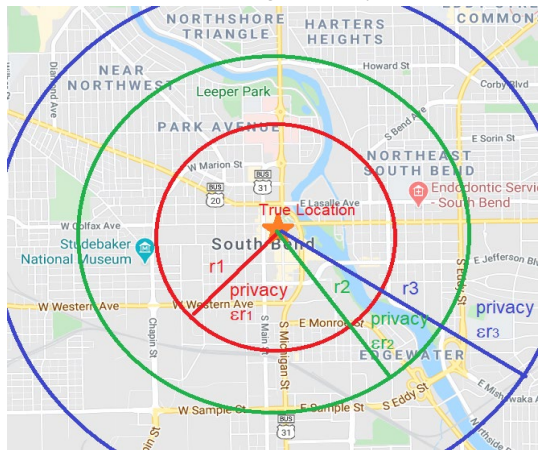

### I.2.3.    Differential Privacy in Practice

Given the attractive mathematical properties of DP, software and web-based interfaces that collect data and release statistics and information with DP have been developed. For example, RescueDP[29] is an online aggregate monitoring scheme that publishes real-time population statistics on spatial-temporal, crowd-sourced data from mobile phone users with DP.[30] PSI ($\Psi$),[31] developed by the Harvard Privacy Tool Project, implements a system for generating and releasing differentially private queries and statistical models, and is integrated with the Dataverse repository,[32] an open source web application to share, cite, explore, and analyze research data.[33] Big companies such as Uber, IBM, and Google have released open-sourced differential privacy libraries at GitHub for experimenting and developing DP applications.[34] The U.S. Census Bureau is pushing the implementation of DP for releasing the 2020 Decennial Census and will extend it to the American Community Survey in the future.[35]

Google, Apple, and other companies have applied local DP to collect users' data. Google employs Randomized Aggregatable Privacy-Preserving Ordinal Response (RAPPOR), an end-user client software to collect Chrome browser data for crowd-sourcing statistics, where local DP is ensured by implementing two types of randomized responses.[36] Figure 3 depicts the system architecture used by Apple for data collection with DP. The system consists of device-side and server-side data processing.[37] The local DP is implemented in two places in this system—when collecting the raw data (the privatization stage) and when the restricted-access server further processes and aggregates data to generate histograms. Uber took a different route and developed an approximate DP method on practical private SQL queries instead of using local DP.

Figure 3: The system architecture for differentially private data collection and processing in Apple, Inc. (modified from the *Apple Machine Learning Journal*[38])

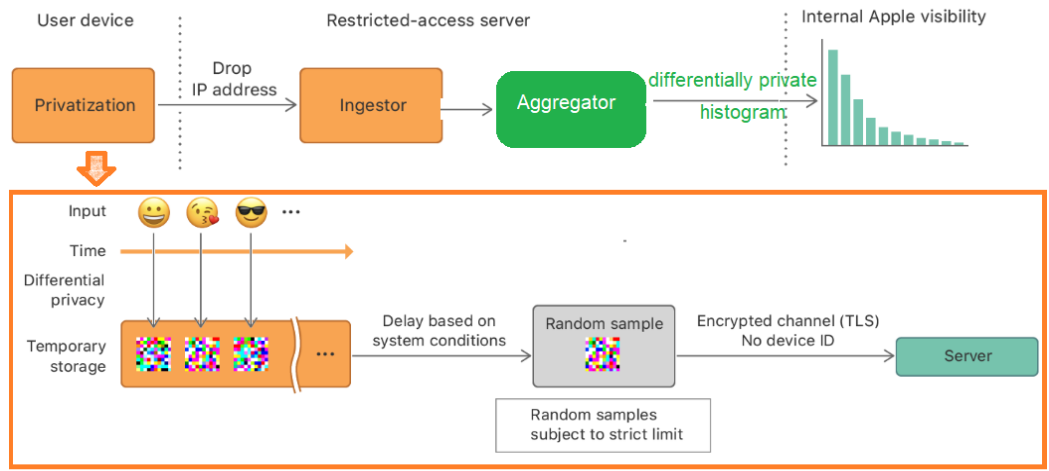

## II. MEASUREMENT OF DATA UTILITY

It is critical to evaluate the utility of the data perturbed for privacy protection reasons before releasing them to the public. If there is too much perturbation, it might render the data useless for research or practical use. To evaluate the balance between privacy protection and data utility, frameworks and approaches for efficiently and accurately assessing utility are needed. In this section, I provide an overview on some existing theoretical and methodological approaches for assessing the data utility. Some approaches focus on a single or a limited set of statistics and query results, while others are more ambitious and aim to assess the overall utility of a released data set.

If the released information contains a finite set of query results or statistics (e.g., histograms, cross-tabulations, means, model parameter estimates, predictions), we may examine the distance between the perturbed statistics and the original statistics, by using the so-called "$l_p$ distance". Some common choices of $p$ is 1, 2, and ∞. When $p = 1$, it becomes the $l_1$ distance (summed absolute differences between the original and perturbed sets); when $p = 2$, it becomes the $l_2$ distance (summed squared differences between the two sets). When $p = ∞$, it becomes the $l_∞$ distance (the maximum absolute distance between two sets). The smaller the distance, the more similar the perturbed set is to the original set, and the more original information are preserved per that metric. Mean squared error (MSE) is another commonly-used metric to assess the utility of the released data. Compared to the $l_p$ distance, MSE captures both the distance and the variability (stability) of the perturbed results.

These utility measures have been employed for utility analysis not only in empirical studies with real-life and simulated data, but also theoretically. Some of the notable theoretical work focuses on bounding the $l_p$ distance, or bounding with a high probability the so-called "excess risk", often defined as the expected $l_p$ distance between the perturbed statistics and the statistics that would be obtained assuming an infinite amount of information. Often times, the bounds are functions of the sample size (or sample complexity), the privacy parameters (such as $\epsilon$ in the DP framework), and some accuracy parameter. With the functional relationships, one can also obtain a lower bound on the sample size given a desired accuracy level with a high probability.[39]

While it is desirable to preserve as much original sample information as possible, the objectives for sharing and releasing data in many cases are to make inferences on the underlying population parameters where the original data are sampled. In others words, if the released data can preserve well the population-level information, then it should be sufficient from a data utility perspective. It has been shown that perturbation, if done appropriately, can help to increase the stability and robustness of the estimates for population parameters, boost the accuracy of predictions, and improve the generalizability of models or parameters learned from training the original data. Therefore, rather than merely focusing on preserving the original sample information, which contains sampling variability and in some cases is harder to achieve, one could instead examine the utility from a statistical inference perspective. For example, in the case of releasing the result for statistical hypothesis testing, rather than focusing on preserving as much as possible the original test statistic value, one may preserve the conclusion for hypothesis testing with a high probability of not violating privacy, which is supposedly an easier task. In the case of releasing parameter estimation (e.g., means or proportions), in addition to releasing a private parameter estimate, one may consider releasing a private uncertainty measurement along with that parameter estimate. If the private confidence interval based on the private point and uncertainty estimates provides close-to-nominal coverage or has a significant overlap with the original confidence interval, then the population-level information might be well-preserved. In summary, shifting the focus to preserving the population-level information will help to alleviate the tension between privacy protection and the data utility as the utility is not about retrieving the individual information in the original data, which is in direct conflict with privacy protection, but rather about preserving population-level information, which is more of an indirect conflict with privacy protection. As a side benefit, the private inferences might offer better generalizability than the inferences based on the original data, which are more susceptible to overfitting. Figure 4 shows a pictorial representation of the idea. The perturbed data deviates from the original sample data. The deviation scenarios in (a) and (b) are acceptable given that the perturbed data still represent the population-level information well, whereas scenario (c) is a bad case where the perturbed data contain biased information about the target population.

**Figure 4**: Relationship between the original data, the perturbed data, and the underlying population where the original data are sampled

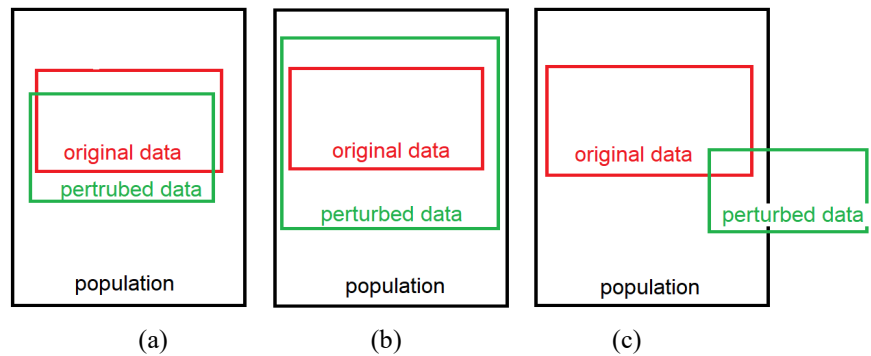

(a)                    (b)                    (c)

The above-discussed utility assessment focuses on assessing the information preservation on a finite set of statistics. Sometimes, the released information is a whole data set, which is often referred to as the "synthetic data" or "surrogate data" (see Section III for more discussion). In this case, it would be nice to have a rather comprehensive metric that shows the general utility of the release data. Strictly speaking, there is no universal data utility measure, unless strong assumptions are made about the distributions of the data. That said, there are some approaches that try to fulfil the goal; and one of the ideas is the propensity-score utility measures, which aim at quantifying the general utility of the synthetic data by measuring their similarity with the original data, regardless the dimensionality of the data. For these approaches, the synthetic data and original data are first combined. A statistical model is then built on the combined data, and the propensity score (the likelihood) of an observation belonging to the synthetic (or the original) data is estimated. There exist several approaches that formulate a single metric based on the property scores that summarizes the similarity between the synthetic and original data.[40] If the synthetic data preserve the original information well, then the observations from the synthetic and original data sets are indistinguishable, and the metric should reflect that. All the approaches focus on preserving the original sample data rather than the population-level information. For example, the synthetic data in scenarios (a) and (b) in Figure 4 are regarded as acceptable from a population-information preservation perspective but might be labelled as "unacceptable" per one of the propensity-score based approaches, especially scenario (b). Further research is warranted and better approaches for evaluating the general utility of the synthetic data are needed.

## III. PERTURBATION METHODS

As mentioned in Sections I and II, *supra*, it is important to find the right balance between sufficient privacy protection and satisfactory utility when developing a data perturbation approach to release information. Different data perturbation approaches will lead to the various relationships between privacy and utility.[41] The ideal case would be maximal utility plus minimal privacy, but this is unachievable in reality; the worst case, minimal utility plus maximal privacy, is unnecessary and defeats the purpose of data release and sharing. Between the two extremes, there is a wide spectrum where data perturbation methods can be explored and developed. The closer the curve is to the ideal case and the larger the area under the curve, the more efficient the corresponding perturbation approach is in protecting privacy yet achieving good utility.

**Figure 5**: Pictorial representation of the privacy versus utility tradeoff. Different solid lines represent different data perturbation approaches for releasing private information

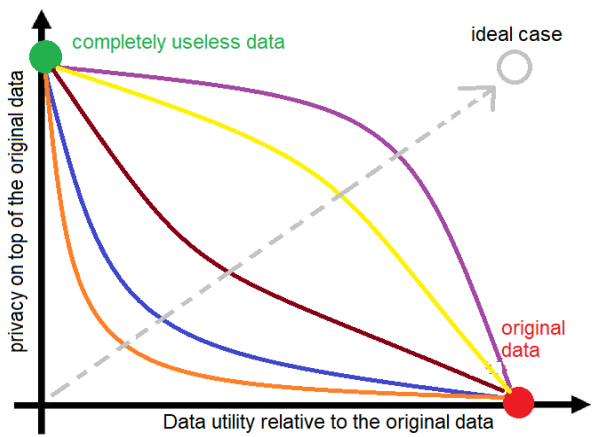

In the rest of this section, I will introduce some existing approaches for data perturbation, including their advantages and disadvantages. Some of the methods are traditional and have existed for as long as almost half a century; others are relatively new and developed within the last decade in the DP framework.

### III.1. Traditional Approaches

Data privacy protection in the statistical community has been known as statistical disclosure limitation or statistical disclosure control (SDL or SDC) for a long time. The methods introduced in this subsection are the traditional approaches for privacy protection before the DP started to dominate the privacy research field. Some of these SDC methods were documented in two major reports from the U.S. Federal Committee on Statistical Methodology in 1978[42] and 1994.[43] Many of these long-existing techniques are still being employed today for releasing data, largely owing to their simplicity and easy implementation. These techniques can be roughly grouped into two categories: model-free and model-based approaches.

Commonly used model-free approaches include global recoding, local suppression, micro-aggregation, data swapping, and post randomization (PRAM). The first three approaches are not random and the information is suppressed in a deterministic way whereas the latter two methods are randomization mechanisms. Global recoding coarsens sensitive information and groups individuals who are at higher risk of being identified and disclosed on their personal information. For example, "annual income," which is often viewed as sensitive personal information, can be coarsened to an ordinal variable (e.g. ≤$31K, $31K–$42K, $42K–$126K, $126K–$188K, >$188K) that is less sensitive. Local suppression removes information that leads to either identification or disclosure of sensitive information in some records. Though global recoding and local suppression provide protection, the incurred information loss can be large and the missing values created by local suppression impose an extra burden on data analysts. Micro-aggregation aims at masking extreme values.[44] It first defines buckets for the values of the attributes, then allocates individual records to these buckets and derives surrogate values for the attributes in each bucket. To reduce information loss, individuals need to be as homogeneous as possible in each bucket in their attributes. Since micro-aggregation acts more on outlying observations, disclosure risk may still exist for sensitive non-outlying cases. Data swapping switches quasi-identifiers[45] among randomly picked individuals[46] or among individuals within the same neighborhood defined based on the similarity of the values on the attribute to be swapped.[47] Data swapping may lead to distorted relationships between swapped and non-swapped attributes. Post Randomization (PRAM) is originally developed to transform values for categorical attributes,[48] and the probabilities of transforming (e.g., from female to male, or the other way around) are stored in the "Markov probability matrix" or "transition matrix". PRAM is performed independently on each individual; and the information loss from PRAM can be large.

Despite the simplicity of the model-free procedures, many of them do not ensure valid statistical inferences based on the perturbed data. Not only are population parameters subject to bias, the uncertainty of these parameter estimates can be under- or over-estimates as well. Model-based approaches can help to solve some of the issues, if applied in a proper way. The model-based approaches often start with building a statistical model on the original data, from which synthetic data (basically predicted values from the model) are generated. The synthetic data might include the whole data set to be released, or only sensitive attributes (such as sexual orientation or financial status) or quasi-identifiers. In the setting of sampling survey, Rubin proposes building a statistical model from the sample data, and predicting non-sampled values of survey variables for the individuals in the population where the sample is drawn.[49] The prediction and sampling can be repeated independently multiple times, and the multiple samples will be released to the public. Little suggests synthesizing a part of the original data, specifically the values of sensitive attributes,[50] which is later extended to the values of quasi-identifiers.[51] To fully account for the uncertainty of the synthetic data, the Bayesian framework is often used where the synthetic data are drawn from the posterior predictive distributions.

If the synthesis model reflects the underlying unknown distribution that generates the original data, then the model-based approaches will lead to valid and more efficient statistical inferences given the synthetic data, compared to the model-free approaches. On the other hand, if the synthesis model deviates from the underlying known distribution that generates the original data, then the synthetic data are not trustworthy. In addition, developing statistical models for high-dimensional data can be very difficult, making it an inefficient approach for dealing with privacy in many situations in the big data era. Since the synthetic data are generated from a statistical model, some perturbation is automatically incorporated with the uncertainty of the synthesis model as well as the sampling error from generating the synthetic data. In addition, the synthetic values are not associated with any real individuals, they are often regarded as providing privacy protection on the individuals who contribute their data in the original data; but a definitive assessment would be needed to quantify the safety level.

### III.2. The Need for New Approaches to Combat Contemporary Privacy Concerns

Advances in analytical techniques and computation, together with the eruption of big data and real-time data collection and processing, can make the traditional methods for privacy protection ineffective. In fact, there have been incidents of database reconstruction and individual identification from supposedly anonymized data. Even if every individual piece of information is stripped of personal information, the relationships between the individual pieces can reveal the individual's identity and a data intruder may still identify an individual in a data set via linkage with other public information. Dinur and Nissim have documented this possibility in their seminal article[52] and proved the "database reconstruction theorem,"[53] also known as the "fundamental law of information recovery."[54] Some notable examples on individual identification breach in publicly released or restricted access data based on database reconstruction include the Netflix prize,[55] the genotype and HapMap linkage effort,[56] and the Washington State health record identification.[57] The U.S. Census Bureau also provides convincing evidence on the need for a more rigorous privacy protection framework in the modern world. Specifically, the Census Bureau conducted an internal experiment to reconstruct and re-identify the 2010 Census records based on the released 150 billion statistics on age, sex, race, ethnicity, and relationship (to householder) for about 309 million individuals.[58] Note that the released statistics are already perturbed using some of the traditional techniques (e.g., global recoding, local suppression, and data swapping). The findings are striking. First, "census block and voting age (18+) were correctly reconstructed for all [309 million] records and for all 6,207,027 inhabited blocks."[59] Second, block, sex, age, race (OMB 63 categories), and ethnicity were reconstructed for 46% of the population (142 million individuals) and within one year for 71% of the population (219 million individuals).[60] Third, applications of record-linkage techniques with external commercial data on block, sex, and age led to putative re-identification of 45% of the population (138 million individuals), and 38% of the identifications (52 million individuals) are correct re-identifications.[61] "The privacy law, in Title 13 of the United States Code, mandates that information about specific individuals, households and businesses is not revealed, even indirectly through our published statistics."[62] And the experiment results from the Census Bureau clearly show that the traditional disclosure avoidance methods are insufficient to counter the privacy risk.

In summary, the new challenges posed by the current paradigm for data collection, analysis, and sharing call for more rigorous privacy concepts and frameworks. Without knowing what intruders know and how powerful they are, privacy protection should aim at dealing with the worst-case scenarios. In other words, the party who releases data should employ a framework that does not make ad-hoc and strong assumptions on how much background information intruders possess, and what algorithms or techniques intruders use to identify individuals, calculate sensitive information, or re-construct a database, given the released data. DP fits the description of such a framework. As mentioned in Section I.2.3, the U.S. Census Bureau is one of the organizations that have decided to adopt DP to guarantee formal privacy to meet its continuing obligations to safeguard respondent information.[63] Specifically, the

---

results from the 2020 census will be published using differentially private mechanisms. As an experiment, it applied the DP mechanism that it will implement (subject to modifications and changes) on the 2020 census to the 2010 census, and the re-identification risk lowered to 0 at privacy budget $\varepsilon = 0$ (as expected), 3% at $\varepsilon = 2$, 4.5% at $\varepsilon = 4$, 6% at $\varepsilon = 8$, and 8.2% at $\varepsilon = 16$. The decreases in the privacy risk are very significant compared to the traditional privacy protection techniques.

### III.3. Differential Privacy Based Approaches

Many approaches and randomization mechanisms have been developed to achieve DP guarantee for released information. Some of them do not target specific statistics or analyses, such as the Laplace mechanism,[64] the Gaussian mechanism for releasing numerical data,[65] and the Exponential mechanism[66] that can release both numerical and categorical data. Differentially private versions of many common statistical analyses and machine learning techniques have also been proposed in the past decade, ranging from simple analyses such as releasing histograms and summary statistics, to common regression models,[67] to principle component analysis,[68] to regularized empirical risk minimization,[69] to deep learning,[70] among others. Releasing one private statistic or a limited set of private results as the interactive query-based data release from the data curator who has access to the original data has drawbacks. Specifically, the requirement to pre-specify the level of privacy budget often dictates the number and the types of future queries with the privacy budget parallel composition theorem,[71] which means that every time the same set of data is queried, there is a privacy cost. Since the overall privacy cost is the sum of all the privacy budgets spent on releasing all queries from that data set, if the sum exceeds the pre-set overall privacy budget, the curator will refuse to answer any further queries. It would be great if data users had direct access to the differentially private individual-level data that are of the same structure as the original data to perform any analysis as if they had the original data.

The DP framework for releasing differentially private individual-level data, is referred to as Differentially Private Data Synthesis (DIPS). Efforts have been made to advance the techniques on DIPS. Some DIPS techniques focus on generating differentially private distributions from which synthetic data can be sampled without imposing assumptions on the distribution of the original data. Other approaches first build a model on the original data, by consuming a portion of the privacy budget, and then generate synthetic data from the constructed model. Though only a fraction of the privacy budget is allocated for data synthesis per se compared to the model-free approaches which use all the budget to generate synthetic data, the benefits of using a well-specified model to generate synthetic data may outweigh the budget splitting between model selection and data synthesis. DIPS approaches to dealing with special types of data have also been developed, such as graphs and networks,[72] and mobility data from GPS trajectories.[73] Interested readers may refer to Bowen and Liu[74] for a more comprehensive overview on DIPS methods.

It is worth noting that the National Institute of Standards and Technology sponsored the Differential Privacy Synthetic Data Challenge in 2018, encouraging development of new methods and improving existing DIPS methods for releasing data, while preserving the dataset's utility for analysis.[75] This competition encourages in-demand work and promotes efforts that help transition from the theoretical work to the practical applications in the area of DP.

## IV. SUMMARY AND DISCUSSION

The world is in the middle of a data and technology revolution. Today's sophisticated computer technology and the increasing data collection and information access have equipped data intruders with more tools to launch privacy attacks successfully. Confidentiality is vital for the future cooperation of individual data contributors and collection of high-quality data to guide policy makers, industries, and businesses to make efficient and timely evidence-based decisions.

Research and practical applications on data privacy have been gaining momentum in recent years. Great progress has been made on the theories and methods in data perturbation, the evaluation of the impact of data perturbation on utility, information loss, statistical inference, and privacy risk measures. In the last decade, DP has been dominating the research on data privacy and there is no sign of slowing down. In addition, practical applications of DP and open-source codes for implementation have started to catch up in recent years. Compared to other parameterized or absolute privacy risk measures, DP does not impose strong assumptions on, or model, data intruder's behavior and external knowledge. As such, it is robust, immune to post-processing, and future-proof. In other words, data intruders cannot gain additional information about his target from released differentially private information no matter what he does to the released information and no matter how much future information he will obtain. These powerful properties of DP and the associated potentials for practical implementation are the key reasons for attracting attention from researchers and practitioners alike.

Big companies such as Google and Apple, and more recently Facebook, have pioneered the implementation of DP techniques in data collection and analysis, the descriptions of which can be easily found online. Online technology news media and blogs (e.g. Wired, AdExchangers, CNET, Science, TechRepublic, Tech Chums) have been closely following the moves and updates regarding the implementation of the DP-related technology in industry and business. Open-source codes have also been made public through GitHub (hosting software development and version control). Despite all the publicity and actual deployment, the DP systems implemented by some of the companies are still subject to conceptual and technical flaws. For example, it is suggested the privacy-loss budget employed by Apple to collect users' data on mobile devices is too high to be acceptable for privacy protection.[76] To gain the trust of consumers who opt to share their data, companies should continually improve their DP systems and models and make sure the privacy risk of their consumers can be controlled at the pre-specified level.

The government agencies, who collect huge amounts of data every year through surveys and who are obligated to share the collected information with taxpayers, should also consider making a transition to apply data perturbation approaches with stricter privacy protection. It is not an easy task to make such a transition, as many government agencies have been using the traditional data perturbation approaches for a very long time. In addition, there are no

---

documented cases and minimal complaints regarding the privacy risk on the information shared by the government agencies over the years, providing few rationales and very weak motivation on the adoption of more formal privacy concept, such as DP. The decision of the U.S. Census Bureau of applying DP in the 2020 Census had occupied some headline news since the official announcement in 2018 and has made many ears prick up. "[C]ritics of the new policy believe the Census Bureau is moving too quickly to fix a system that isn't broken. They also fear the changes will degrade the quality of the information used by thousands of researchers, businesses, and government agencies."[77] It is argued that database reconstruction, similar to what the Census Bureau has experimented with, have been exaggerated, and the DP approach is inconsistent with the statutory obligations, history, and core mission of the Census Bureau.[78] I would argue that the DP approach adopted by the Census Bureau might seem like overprotection for the present, but the privacy protection offered by DP covers the worst-case scenario and is future-proof. This is very important as no one can predict what the future is like for certain. In addition, if the DP strategies adopted by the Census Bureau on the 2020 census turn out to yield useless data (which is very unlikely) by injecting too perturbation for privacy purposes, the Census Bureau could always fall back on the traditional ways of releasing data, without losing much in either privacy or utility, except for the resources or manpower devoted to the DP project.

Despite the rapid growth in the field of DP, issues and opportunities in research and applications exist. The field of DP has witnessed a huge jump in the number of manuscripts in recent years. Papers on DP were published quickly due to its popularity at the expense of quality. For example, there have been cases where a published manuscript claims satisfaction of a certain $\epsilon$-level, but the actual privacy cost is much higher; or, two manuscripts discuss the same topic, but draw conflicting conclusions.[79] Peer-reviewed journals and conferences will need to better police the research on data privacy, especially on DP.

I list below some potential areas in DP and data privacy in general that will benefit from more work and further research. First, more work is needed to test the scalability and feasibility of the developed DP methods in practice. Though many published DP techniques share empirical evidence on their potentials for practical implementation by running experiments and tests on publicly available data, many of the employed data are a "nice" subset with significantly fewer attributes than the raw data, or pre-processed by removing hard-to-deal problems, such as cases with missing values. The "real" real-life data can be much nastier and messier than those used by researchers in their manuscripts. Not only are there missing values, data entry errors, and outliers, but also data might exhibit correlation and the dimensionality can be huge (number of attributes in the hundreds or thousands). In the case of survey data, complex survey designs and weighting schemes might be used, which cannot be ignored when performing analysis. Improvement and extension of available approaches and development of new methods should be able to handle these real-life challenges as much as possible. Second, easy-to-use, user-friendly, and computationally-cheap software and tools to implement data perturbation with a guarantee of formal privacy and to measure data utility are still in great need. Without proper tools developed and deployed in time to facilitate application, the excitement about a new approach or technique will soon disappear and be abandoned by practitioners quickly. Third, most research on privacy in computer science focuses on prediction and might not be easily extendable to private statistical inferences. Statisticians and computer scientists should work together to provide effective solutions to this problem. Fourth, if a parameterized privacy protection technique is used, choosing appropriate hyper-parameters (both privacy and accuracy parameters) is critical. In the case of DP, there is still no consensus on the appropriate amount of the privacy budget for practical use. Even with the 2020 Census just around the corner, the Census Bureau is still searching for a suitable privacy budget for the 2020 census. Finally, there is an urgent need to educate the public on what DP is, especially if it will be widely used for collecting and releasing data in the near future. Efforts have been made. For example, the YouTube video "Protecting Privacy with MATH," a collaboration between minute physics and the Census Bureau, posted on September 12, 2019, has had 334,290 views as of March 9, 2020.[80] The American Statistical Association Committee on Privacy and Confidentiality, which I am part of, regularly invites speakers to talk about DP through webinars on Data Privacy day, which offers free registration to the public. The fact that household-name big companies employ DP in data collection and release and that the Census Bureau will use DP for releasing 2020 Census data have helped grab some people's attention to DP. Despite all these efforts, there are still confusions about what formal privacy means, not only among the public, but also among researchers who are not specialized in privacy research.  Industries, government agencies, and researchers who are adopting DP or conducting research in DP should make more efforts to reach out to the public on the increase in privacy risk that is largely due to the collection of big data and recent technical advances, the state-of-art privacy research, the techniques that can help combat the new privacy threats, and the importance and necessity of continuing data collection and sharing under the formal privacy guarantee.

In summary, to help people to exercise their right to privacy, policy makers, legislators, industries and businesses, academia and research institutes, and individuals/consumers will need to work together to achieve that goal.  It is not an easy task, but something has to be done not only for ourselves, but also for many generations to come.